# Diverse ATPase proteins in mobilomes constitute a large potential sink for prokaryotic host ATP

Hyunjin Shim[1], Haridha Shivram[2], Shufei Lei[1], Jennifer A. Doudna[2], Jillian F. Banfield[1,2,3,4,5,6†]

[1]Earth and Planetary Science, University of California, Berkeley, CA, USA
[2]Innovative Genomics Institute, University of California, Berkeley, CA, USA
[3]Department of Environmental Science, Policy, and Management, University of California, Berkeley, CA, USA
[4]Earth Sciences Division, Lawrence Berkeley National Laboratory, Berkeley, CA, USA
[5]Chan Zuckerberg Biohub, San Francisco, CA, USA
[6]University of Melbourne, Melbourne, VIC, 3010, Australia

†Corresponding Author: jbanfield@berkeley.edu

## Abstract

Prokaryote mobilome genomes rely on host machineries for survival and replication. Given that mobile genetic elements (MGEs) derive their energy from host cells, we investigated the diversity of ATP-utilizing proteins in MGE genomes to determine whether they might be associated with proteins that could suppress related host proteins that consume host energy. A comprehensive search of 353 huge phage genomes revealed that up to 9% of the proteins have ATPase domains. For example, ATPase proteins constitute ~3% of the genomes of Lak phages with ~550 kbp genomes that occur in the microbiomes of humans and other animals. Statistical analysis shows the number of ATPase proteins increases linearly with genome length, consistent with a large sink for host ATP during replication of megaphages. Using metagenomic data from diverse environments, we found 505 mobilome proteins with ATPase domains fused to diverse functional domains. Among these composite ATPase proteins, 61.6% have known functional domains that could contribute to host energy diversion during the mobilome life cycle. As many have domains that are known to interact with nucleic acids and proteins, we infer that numerous ATPase proteins are used during replication and for protection from host immune systems. We found a set of uncharacterized ATPase proteins with nuclease and protease activities, displaying unique domain architectures that are energy intensive based on the presence of multiple ATPase domains. In many cases, these composite ATPase proteins genomically co-localize with small proteins in genomic contexts that are reminiscent of toxin-antitoxin systems and phage helicase-antibacterial helicase systems. Small proteins that function as inhibitors may be a common strategy for control of cellular processes, thus could inspire the development of new nucleic acid and protein manipulation tools, with diverse biotechnological applications.

# Introduction

The mobilome in prokaryotes includes plasmids, bacteriophages (phages), viruses of archaea and transposons that can move between host genomes. These mobile genetic elements are dependent on host machineries for transcription and replication [1–3]. This parasitic propagation strategy necessitates the diversion of host energy, as these mobile genetic elements lack energy generating machinery (e.g., ATP synthase) and structures (e.g., the cell membrane). The life cycle of a phage requires several energy intensive steps, both in the lytic and lysogenic cycles (Figure 1) [4]. In the lytic cycle, phages have a DNA or RNA genome that is encapsulated within particles that are released and go on to infect new cells. In the lysogenic cycle, the phage can integrate into the host genome. These intra-genomic conflicts between hosts and mobilomes constitute a huge ATP sink.

Biochemically, protein activities that involve conformation changes in the structure require energy processing modules such as ATPase domains to bind and hydrolyze ATP to ADP. For example, ATP-dependent proteins such as helicases use energy released from ATP hydrolysis to unwind DNA segments during replication [5,6]. At this time, the mobile genetic elements use host machinery either to replicate their genomes prior to cell lysis or to integrate into the host genome. Machineries involved in replication that are ATP-dependent include nucleases, transposases and polymerases [7–11]. Further, the capsid packaging step is energetically expensive, approximately one ATP is required by the terminase to package two base pairs into the capsid for active DNA packaging [12].

Prokaryotes have evolved several resistance mechanisms that are often ATP-dependent processes to disable the invasion of foreign genetic elements, notably restriction-modification systems [10] and some CRISPR-Cas systems [13]. Restriction modification systems consist of restriction endonucleases that cleave and degrade specific DNA sequences that are not modified by methylases [14–16]. Studies have shown restriction modification systems (including restriction-only or modification-only systems) are energetically expensive [17,18]. For example, it was estimated that *E. coli* could consume ~0.2% of its total ATP pool for effective restriction against invading foreign genomes [17]. CRISPR-Cas systems are an adaptive immune system that can acquire and degrade foreign genetic elements [19–22]. Recent discovery of transposon-encoded CRISPR-Cas variants Tn6677 of *Vibrio cholerae* in *Escherichia coli* reveals that the system requires an ATPase protein for transposition [23–25].

As a counter defense, some phages possess small anti-restriction proteins that impede or inhibit the activities of restriction enzymes [26–29]. A recent study revealed the phage-encoded small protein that inhibits the ATPase domain of the *Staphylococcus aureus* helicase loader [5]. This could allow the phage to substitute its own helicase but it also prevents ATP consumption by the bacterial helicase, conserving ATP for phage functions. Even more recently, it was shown that tiny phage-encoded hypothetical proteins block bacterial quorum sensing machinery. From these few cases, and given their substantial energy requirements, we deduce that phages probably have a variety of proteins that have evolved to block bacterial systems that consume ATP.

Here, we sought to identify ATPase proteins in the genomes of mobile genetic elements (mobilome) assembled from metagenomic sampling. This bioinformatic analysis used a dataset of huge phage genomes to survey the relative abundance of ATP-requiring proteins and a large metagenomic dataset from various environments to discover new types of ATPase proteins in mobilome genomes. Particularly, we focus on proteins containing domains involved in nucleic acid and protein interactions, with the long-term aim of discovering new potential candidates for nucleic acid or protein manipulation tools. Furthermore, we investigated small proteins co-localized and co-occurring with ATP-requiring proteins to generate a database of protein candidates that may inhibit host ATPases, enabling mobile elements to hijack bacterial energy resources.

## Results

**ATPase genes constitute up to 9% of the huge phage genomes.**

The ATPase proteins predicted from huge phage genomes account for, on average, ~3% of the proteins (average 12.59 per genome have at least one ATPase domain; Table 1). Some of these ATPase proteins have multiple ATPase domains (1.69 per genome, 14.29ATPase domains per genome). The variance in the number of ATPase proteins is wide, as shown as the standard deviation in Table 1, with one phage genome of 706 kbp (M01_PHAGE_CU_48_59) containing 47 ATPase proteins, whereas another phage genome of 220 kbp (FFC_PHAGE_43_1208) contains only 4 ATPase proteins. In one genome, ATPase proteins constitute 9% of the gene content (F31_PHAGE_39_18, 336 kbp). As the phage genome size varies from 104 kbp to 735 kbp, the number of ATPase proteins was plotted against the genome size (Figure S1A). The regression shows a linear relationship between genome length (p-value = $2 \times 10^{-16}$) and the number of ATPase proteins (4.435 per 100 kb).

As an example, we used a well sampled clade of Lak phages of ~550 kbp genomes to conduct a detailed analysis of ATPase proteins in a specific clade. Lak phages are currently the largest genomes identified in human and animal microbiomes [30,31]. As shown in Figure 2, the predicted ATPase proteins are distributed throughout the genome and have diverse functions, including topoisomerase, transferase, protease, kinase, ligase, and helicase, while at least 7 of them have unknown functional domains or no functional domains.

**ATPase proteins are diverse and numerous in mobilome genomes.**

To further analyze the function of ATPase proteins in mobilomes, we searched for composite ATPase proteins that have an ATPase domain fused with one or more functional domains. From the mobilome dataset, we found 505 composite proteins that had an ATPase domain fused with at least one functional domain. These were categorized by analyzing the annotations on the function domains. Surprisingly, 194 of the 505 composite ATPase proteins (38.4%) have domains with no match to functional annotations.

The other 311 composite ATPase proteins (61.6%) had domains with annotations related to diverse functions associated with various stages of the mobilome life cycle, including nucleic acid and amino acid interacting enzymes, and packaging and transport related proteins. Among these, we found 90 composite ATPase proteins with annotations that are related to later stages of the mobilome life cycle such as packaging, transport and secretion. The remaining 221 composite

ATPase proteins are nucleic acid and amino acid interacting enzymes, and thus are potentially involved in replication and defense. As shown in Figure 3, this category includes nucleic acid interacting proteins such as nucleases and helicases, and amino acid interacting proteins such as proteases. The most abundant functions of these composite ATPase proteins are proteases, nucleases, helicases and DnaK chaperones. Other annotations indicate functions related to membrane-bound proteins for transport and secretion, which have recently been suggested to be associated with defense systems [32,33].

**Some ATPases involved in replication are not yet functionally characterized.**
We found several composite ATPase proteins with domains of unknown function (DUF) or uncharacterized function. These proteins are annotated in the search database with some superfamily domains based on sequence or structural similarity, but the exact biological function of each family has not been characterized. For instance, a few composite ATPase proteins have PIN domains that belong to a superfamily of nucleases with distinct structural similarities of a common Rossmanoid fold [34,35]. They are known to be broadly involved in central cellular processes, but each family participates in diverse central cellular processes such as DNA replication and repair, transcription regulation, mRNA degradation and ncRNA maturation. As shown in Figure S2, the PIN-like domains found in these composite ATPase proteins belong to the PIN_4 family. The PIN_4 family was reported to occur in bacteria and eukaryotes, but the exact function is unknown [35]. Interestingly, two of these proteins also have another nuclease superfamily of the LAGLIDADG homing endonucleases fused as a domain (Figure S2). These 'meganucleases' are DNA cleaving enzymes in microbial genomes and cut target sites with high specificities to initiate the lateral transfer of mobile genetic elements [36].

Another category of interest has domains that are related to the virulence-associated protein E (virE), which is known to be part of a prophage region that is associated with virulence in several microbes [37,38]. However, the exact mechanism by which the virE affects virulence has not been characterized. Other domains of interest include those involved in DNA repair (e.g., MutS) and nucleotide binding (e.g., CobQ/CobB/MinD/ParA domain).

**ATPase domains cluster by protein function in multiple sequence alignments and in tree phylogenies.**

To visualize the patterns, the ATPase domains of the composite ATPase proteins were aligned by sequence similarity (Figure 4A). Figure 4A shows the ATPase domains are a 300 amino acid sequences of high variability, with some conserved motifs, including the ATP binding motif (Walker-A) and the ATP hydrolysis motif (Walker-B). The Walker-A motif in the ATPase domains has a primary amino acid sequence of GxxGxGKT or GxxGxGKS, where the letter x represents any amino acid [39]. The Walker-B motif has a primary sequence of hhhhD, where the letter h represents any hydrophobic amino acid (including Valine, Isoleucine, Leucine, Phenylalanine, Methionine). Structurally, the Walker-A motif is an α-helix followed by a glycine-rich loop, and the Walker-B motif is a β-strand connected to the Walker-A by a 100-peptide sequence.

The multiple sequence alignment shows that the ATPase domains cluster by the biological or enzymatic activity of the functional domain. The upper clusters tend to be the biological activities related to the later stages of infection, such as assembly and lysis, including chaperones and transport proteins. The lower clusters tend to be the enzymatic activities related to the earlier stages of infection, such as replication and defense, including nuclease, polymerase, protease and helicase.

Using the multiple sequence alignment from Figure 4A, a phylogenetic tree of the ATPase domains was reconstructed (Figure 4B). The tree shows that the ATPase domains cluster by the predicted function. Interestingly, the family of an ATPase domain and the predicted function of its associated domain classified according to the secondary structure [7–10] closely aligns with the predicted ATPase function. For example, most ATPase domains associated with transport belong to the ABC family, while many ATPase domains associated with helicases and polymerases belong to the helicase superfamily [10]. The helicase superfamily consists of one active ATPase domain fused with a lid that is an inactive ATPase domain. Furthermore, most ATPase domains of unknown function belong to the AAA+ family that are associated with diverse cellular activities [7].

**Genomic context of composite ATPase proteins is conserved.**

The analysis of protein lengths and genomic context shows that the composite ATPase proteins are often co-localized with small proteins (Figure S3). It is notable that the average length of the composite ATPase proteins is more than double than the average length of the co-localized proteins. To identify if these small proteins are conserved, homologs for each protein in the genomic context of three proteins downstream and upstream of the composite ATPase proteins were identified (Figure S4). As shown in Figure 5A, the proteins in the genomic context of the composite ATPase proteins have many close homologs. Hierarchical clustering reveals that genomic contexts have three main types: the first with many homologs across the genomic context (bottom), the second with many homologs in the proteins adjacent to the composite ATPase proteins (top), and the third with few homologs in the genomic context (middle).

      The first and the second clusters were investigated further for a conserved genomic context (Table S1). The analyzes reveal that the order of protein homologs was better conserved when the protein has more homologs, which is consistent with the hypothesis that the two factors are correlated. Some genomic contexts of the composite ATPase proteins from the first cluster and the second cluster are shown in Figures 5B and 5C, respectively. It is notable that most of the co-localized proteins are hypothetical, but yet conserved in this genomic context. Figure 5B shows the highly conserved genomic context of the composite ATPase proteins, whose function is mainly related to proteolytic activities. These composite ATPase proteins with multiple ATPase domains belong to the family of Lak megaphages. Figure 5C shows the genomic contexts with many homologs in the proteins adjacent to the composite ATPase proteins. Interestingly, these composite ATPase proteins display diverse functions, such as nuclease, protease, terminase, primase and helicase, and they overlap in sequence with the adjacent proteins. These genes that overlap in sequence with the composite ATPase proteins are more conserved than non-overlapping genes in the genomic context.

**Discussion**

To survive and replicate, mobile genetic elements require ATP available within the host cell. For each step of the mobilome life cycle, we identified many ATPase proteins that are potentially involved in energy-requiring activities. A comprehensive survey of 353 huge phage genomes revealed that, on average, 3% of the proteins are ATP-binding, although some genomes encode up to 9% ATP-binding proteins. Given that statistical analysis shows the number of ATPase proteins increases linearly with the genome length, phages with large genomes represent a huge sink for bacterial ATP. Lak megaphages, whose genomes are >20% of the length of their host *Prevotella* genomes, have diverse ATPase proteins, including topoisomerases, transferases, proteases, kinases, ligases, helicases, and proteins of unknown functions. Thus, these phages invest substantial host energy in the form of ATP in nucleic acid, as well as protein transformations.

We analyzed the functions of ATPase proteins in mobile genetic elements from diverse environments. Of the 505 composite ATPase proteins with additional domains, 38.4% had domains of unknown function. A small subset of membrane-bound proteins appears to be involved in defense [32,33]. Most of the remainder of the composite ATPase proteins have functional domains that suggest roles during the early stages of infection or replication, such as nuclease, protease, polymerase, helicase and transposase. Consistent with these indications of energetically expensive stages of the mobilome activity, we discovered some composite ATPase proteins with up to three ATPase domains per protein. These proteins are mostly predicted to be nucleases, proteases, polymerases and chaperones. Some of the most interesting proteins in this category are predicted to be site-specific nucleases of the PIN domain and LAGLIDADG domain fused with multiple ATPase domains. The explanation for nucleases that require multiple ATPase domains is not known. These may be candidates of interest for experimental evaluation.

Mobilome genomes encode large inventories of small proteins, the functions of which remain a mystery. We analyzed the genomic contexts of some small proteins encoded adjacent to ATPase domain proteins, with the objective of developing hypotheses regarding their roles. We discovered genes for several small hypothetical proteins that overlap with genes encoding composite ATPase proteins. This colocalization pattern is conserved in many examples. This is reminiscent of the genomic architecture of a previously validated system in which the ATPase is a toxin and the small protein inhibitor an antitoxin. Another study found an inhibitory role for a small phage protein, the gene for which overlapped with a gene for an ATPase protein [5,6]. Specifically, a phage-encoded small protein was shown to inhibit the ATPase domain of the *Staphylococcus aureus* helicase loader [5]. Based on the combination of our results and these findings, we suspect that small proteins associated with ATP-binding proteins may inhibit host homologs and substitute for their function, simultaneously eliminating an undesirable sink for host ATP. In fact, we speculate that many phage-borne small proteins have evolved to block host bacterial machinery, a concept that is supported by the discovery of small anti-CRISPR proteins that protect phages by inhibiting the function of CRISPR-Cas bacterial immune systems (Figure S5). An additional indication that small proteins are part of a widespread phage strategy to control host cells is the very recent discovery that small phage-encoded proteins bind to and block the regulator of the bacterial quorum sensing pathway. These proteins also bind to the pilus assembly ATPase protein to prevent superinfection by other phages [40].

It is known from experimental studies that phage-encoded ATPase inhibitors bind specifically the host ATPase proteins and not the phage ATPase homolog [5,6]. Thus, an interesting question relates to why small proteins block the ATP binding sites of the bacterial protein but not that of the mobilome homolog. We found that the ATPase domains in the mobilome display highly varying degrees of sequence conservation. All have functionally critical motifs (Walker-A and Walker-B) and variations in other parts of the sequences, including Motif-C and other sensing regions. We predict that small proteins that bind and inhibit the ATP binding site interact with these more variable regions. This would enable the specificity required to avoid blocking of the phage protein. If correct, this will open new opportunities to discover and design small protein inhibitors, with broad applications in areas such as antimicrobial therapy and cancer-suppression.

## Materials and Methods

### ATPase domain predication in the huge phage dataset

The huge phage dataset contains 353 near complete genomes from diverse environments, including rivers, lakes, groundwater, sea, soil and human/animal microbiomes [41]. The dataset consists of 140,206 proteins. The genes of some huge phages such as Lak phages were predicted using Prodigal and genetic code 15, as the stop codon (TAG) has been reprogrammed to encode glutamine [31]. ATPase domains from the huge phage dataset were predicted with the e-value cut-off of $1 \times 10^{-3}$ based on the accession of their Hmmsearch match against the Pfam database (version 33.1) [42]. The Pfam domain search was compared and supplemented with the Hmmsearch match against the Supfam database for structural information [43]. The number of ATPase proteins was plotted against the phage genome size (Figure S1A). A linear regression was fitted after the two distributions were evaluated with the Normal Q-Q plot to check they are linearly related and normally distributed (Figure S1B).

### Big mobilome dataset and curation

The initial mobilome dataset contains 91,205 contigs from ggKbase that were identified based on their gene content (last accessed in August 2019). ggKbase is a web application designed to assist in aggregating, visualizing, exploring and analyzing metagenomic data. The dataset encompasses 1,309,155 proteins in total; 1,052,230 phage proteins, 44,650 virus proteins, 16,467 plasmid proteins; and 35,695 others which includes proteins from the contigs that were manually binned as mobilome (Figure S6). This dataset was used to investigate ATP-requiring proteins. Each protein was first annotated using the basic local alignment search tool (BLAST: version 2.10.1) against the NCBI database. The initial search for ATPase proteins was done by searching for annotations of the mobilome proteins with the keyword 'ATPase' (Figure S7). The mobilome dataset had 5,309 proteins with the matching criteria.

### Composite gene families detection

Selecting composite ATPase proteins from the mobilome dataset was achieved using CompositeSearch with a two-step procedure. First, the initial ATPase dataset from the mobilome dataset was used to find homologs using BLAST against the NCBI RefSeq non-redundant database and the mobilome dataset. The resulting ATPase dataset contained 1,433,097 ATPase protein homologs. Second, this homolog dataset was used to search for composite ATPase proteins in the NCBI RefSeq non-redundant database and the mobilome dataset, where the sequence similarity networks was used to detect composite gene families. The output file from CompositeSearch contained 64,089 composite ATPase proteins in total. Only 1,003 composite ATPase proteins from the ggKbase metagenomic dataset were selected for further analyzes.

Among this composite ATPase proteins from the mobilome dataset, some were from contigs that had bacterial or archaeal taxonomic profiles. To eliminate contigs that are likely to be non-mobilome, the composite ATPase proteins from the contigs that were identified to encode >50% proteins with best hits to bacteria or/and archaea were excluded from further analysis, resulting in a set of 562 candidates. Finally, after eliminating the 57 duplicate candidates generated by the gene prediction program (Prodigal version 2.6.3), the final set of composite ATPase proteins was 505.

**Functional annotation and domain prediction**

Composite ATPase proteins were functionally annotated based on the protein signature recognition of their best InterProScan match (version 5.47-87) [44]. InterPro Application Programming Interface (API) was used for direct access to the InterPro database (version 81.0), with the e-value cut-off of $1 \times 10^{-3}$. Domains were predicted based on the accession of their best Hmmsearch match against the Pfam database (version 33.1), using the same procedure. The composite ATPase proteins with the functional domains of interest were further explored using TRee-based exploration of neighborhoods and domains (TREND) [45]. This platform uses MAFFT algorithms for multiple protein sequence alignment (-auto option) and uses FastTree (version 2.0.0) [46] to perform tree reconstruction. The resulting tree-based exploration of domains can be visualized with the domain architecture identified from the various sources, including Pfam, CDD, NCBI and MiST.

**Multiple sequence alignment of predicted ATPase domains**

ATPase domains were extracted using the domain predictions from the Pfam database (version 33.1) by retaining only the domains with the annotation keywords "ATP" or "AAA". Protein sequences of the ATPase domains were extracted through the start and end predictions of the domain for each composite ATPase protein. The resulting ATPase domains had 179 sequences, which were aligned using a multiple sequence alignment program, MAFFT (version 7.471) (-auto option) [47]. Subsequently, the multiple sequence alignment of the predicted ATPase domains was visualized using Jalview (version 2.11.1.3). The residue positions are coded with the Taylor protein coloring scheme [48] using the conservation visibility of 15%.

**Phylogenetic analyses of predicted ATPase domains**

Tree reconstructions from the multiple sequence alignments of the predicted ATPase domains were performed using IQ-TREE (version 1.6.12) [49]. From the ModelFinder [50], the best performing substitution model was selected as LG+R5, and 1000 ultrafast bootstrap were run [51]. The reconstructed tree was visualized with iTOL (version 4) and iTOL annotation editor [52], with the following labels: protein name with predicted function, ATPase classification and sample location.

**Identification of genomic context conservation near composite ATPase proteins**

The genomic context of the composite ATPase proteins was defined as proteins co-localized three genes upstream and downstream. First, each protein within the genomic context was searched for homologs in the ggKbase database using BLAST [53] with the e-value cut-off of $1 \times 10^{-30}$. The BLAST result for each composite ATPase protein and its neighboring proteins was summarized as a heatmap. The intensity of color represents the number of BLAST hits, which is equivalent to the number of homologs in the ggKbase. The heatmap was clustered hierarchically by similarity for pattern recognition. The BLAST hits were analyzed for conservation of order in different clusters. If the order of the genomic context was conserved, these ATPase proteins were defined to have a conserved genomic context that may be functionally linked as an operon.

# References

1. Frost, L. S., Leplae, R., Summers, A. O. & Toussaint, A. Mobile genetic elements: The agents of open source evolution. *Nat. Rev. Microbiol.* **3**, 722–732 (2005).
2. Gogarten, J. P. & Townsend, J. P. Horizontal gene transfer, genome innovation and evolution. *Nat. Rev. Microbiol.* **3**, 679–687 (2005).
3. Thomas, C. M. & Nielsen, K. M. Mechanisms of, and barriers to, horizontal gene transfer between bacteria. *Nat. Rev. Microbiol.* **3**, 711–721 (2005).
4. Knowles, B. *et al.* Lytic to temperate switching of viral communities. *Nature* **531**, 466–470 (2016).
5. Hood, I. V. & Berger, J. M. Viral hijacking of a replicative helicase loader and its implications for helicase loading control and phage replication. *Elife* **5**, (2016).
6. Liu, J. *et al.* Antimicrobial drug discovery through bacteriophage genomics. *Nat. Biotechnol.* **22**, 185–191 (2004).
7. Neuwald, A. F., Aravind, L., Spouge, J. L. & Koonin, E. V. AAA+: A class of chaperone-like ATPases associated with the assembly, operation, and disassembly of protein complexes. *Genome Res.* **9**, 27–43 (1999).
8. Iyer, L. M., Leipe, D. D., Koonin, E. V. & Aravind, L. Evolutionary history and higher order classification of AAA+ ATPases. in *Journal of Structural Biology* **146**, 11–31 (2004).
9. Castaing, J. P., Nagy, A., Anantharaman, V., Aravind, L. & Ramamurthi, K. S. ATP hydrolysis by a domain related to translation factor GTPases drives polymerization of a static bacterial morphogenetic protein. *Proc. Natl. Acad. Sci. U. S. A.* **110**, E151–E160 (2013).
10. Krishnan, A., Burroughs, A. M., Iyer, L. M. & Aravind, L. Comprehensive classification of ABC ATPases and their functional radiation in nucleoprotein dynamics and biological conflict systems. *Nucleic acids research* **48**, 10045–10075 (2020).
11. Shim, H. Feature Learning of Virus Genome Evolution With the Nucleotide Skip-Gram Neural Network. *Evol. Bioinforma.* **15**, 117693431882107 (2019).
12. Yang, Q. & Catalano, C. E. Biochemical characterization of bacteriophage lambda genome packaging in vitro. *Virology* **305**, 276–287 (2003).
13. Sinkunas, T. *et al.* Cas3 is a single-stranded DNA nuclease and ATP-dependent helicase in the CRISPR/Cas immune system. *EMBO J.* **30**, 1335–1342 (2011).
14. Loenen, W. A. M. & Raleigh, E. A. The other face of restriction: Modification-dependent enzymes. *Nucleic Acids Research* **42**, 56–69 (2014).
15. Ershova, A. S., Rusinov, I. S., Spirin, S. A., Karyagina, A. S. & Alexeevski, A. V. Role of restriction-modification systems in prokaryotic evolution and ecology. *Biochem.* **80**, 1373–1386 (2015).
16. Vasu, K. & Nagaraja, V. Diverse Functions of Restriction-Modification Systems in Addition to Cellular Defense. *Microbiol. Mol. Biol. Rev.* **77**, 53–72 (2013).
17. Roberts, G. A. *et al.* An investigation of the structural requirements for ATP hydrolysis and DNA cleavage by the EcoKI Type i DNA restriction and modification enzyme. *Nucleic Acids Res.* **39**, 7667–7676 (2011).
18. Seidel, R., Bloom, J. G. P., Dekker, C. & Szczelkun, M. D. Motor step size and ATP coupling efficiency of the dsDNA translocase EcoR124I. *EMBO J.* **27**, 1388–1398 (2008).
19. Barrangou, R. *et al.* CRISPR provides acquired resistance against viruses in prokaryotes. *Science (80-. ).* **315**, 1709–1712 (2007).

20. Tyson, G. W. & Banfield, J. F. Rapidly evolving CRISPRs implicated in acquired resistance of microorganisms to viruses. *Environ. Microbiol.* **10**, 200–207 (2008).
21. Makarova, K. S., Grishin, N. V., Shabalina, S. A., Wolf, Y. I. & Koonin, E. V. A putative RNA-interference-based immune system in prokaryotes: Computational analysis of the predicted enzymatic machinery, functional analogies with eukaryotic RNAi, and hypothetical mechanisms of action. *Biology Direct* **1**, 7 (2006).
22. Makarova, K. S. *et al.* Evolutionary classification of CRISPR–Cas systems: a burst of class 2 and derived variants. *Nat. Rev. Microbiol.* (2019). doi:10.1038/s41579-019-0299-x
23. Klompe, S. E., Vo, P. L. H., Halpin-Healy, T. S. & Sternberg, S. H. Transposon-encoded CRISPR–Cas systems direct RNA-guided DNA integration. *Nature* **571**, 219–225 (2019).
24. Peters, J. E., Makarova, K. S., Shmakov, S. & Koonin, E. V. Recruitment of CRISPR-Cas systems by Tn7-like transposons. *Proc. Natl. Acad. Sci. U. S. A.* **114**, E7358–E7366 (2017).
25. Strecker, J. *et al.* RNA-guided DNA insertion with CRISPR-associated transposases. *Science* eaax9181 (2019). doi:10.1126/science.aax9181
26. Iida, S., Streiff, M. B., Bickle, T. A. & Arber, W. Two DNA antirestriction systems of bacteriophage P1, darA, and darB: characterization of darA- phages. *Virology* **157**, 156–166 (1987).
27. Kelleher, J. E. & Raleigh, E. A. Response to UV damage by four Escherichia coli K-12 restriction systems. *J. Bacteriol.* **176**, 5888–5896 (1994).
28. Tock, M. R. & Dryden, D. T. F. The biology of restriction and anti-restriction. *Curr. Opin. Microbiol.* **8**, 466–472 (2005).
29. Wilkins, B. M. Plasmid promiscuity: Meeting the challenge of DNA immigration control. *Environ. Microbiol.* **4**, 495–500 (2002).
30. Crisci, M. A. *et al.* Wide distribution of alternatively coded Lak megaphages in animal microbiomes. *bioRxiv* 2021.01.08.425732 (2021). doi:10.1101/2021.01.08.425732
31. Devoto, A. E. *et al.* Megaphages infect Prevotella and variants are widespread in gut microbiomes. *Nat. Microbiol.* **4**, 693–700 (2019).
32. Doron, S. *et al.* Systematic discovery of antiphage defense systems in the microbial pangenome. *Science (80-. ).* **359**, (2018).
33. Millman, A., Melamed, S., Amitai, G. & Sorek, R. Diversity and classification of cyclic-oligonucleotide-based anti-phage signalling systems. *Nat. Microbiol.* 1–8 (2020). doi:10.1038/s41564-020-0777-y
34. Arcus, V. L., Mckenzie, J. L., Robson, J. & Cook, G. M. The PIN-domain ribonucleases and the prokaryotic VapBC toxin-antitoxin array. *Protein Engineering, Design and Selection* **24**, 33–40 (2011).
35. Matelska, D., Steczkiewicz, K. & Ginalski, K. Comprehensive classification of the PIN domain-like superfamily. *Nucleic Acids Research* **45**, 6995–7020 (2017).
36. Taylor, G. K. *et al.* LAHEDES: The LAGLIDADG homing endonuclease database and engineering server. *Nucleic Acids Res.* **40**, W110 (2012).
37. Ji, X. *et al.* A novel virulence-associated protein, vapE, in Streptococcus suis serotype 2. *Mol. Med. Rep.* **13**, 2871–2877 (2016).
38. Ma, Z. *et al.* Insight into the specific virulence related genes and toxin-antitoxin virulent pathogenicity islands in swine streptococcosis pathogen Streptococcus equi ssp. zooepidemicus strain ATCC35246. *BMC Genomics* **14**, 377 (2013).
39. Schneider, E. & Hunke, S. ATP-binding-cassette (ABC) transport systems: Functional and structural aspects of the ATP-hydrolyzing subunits/domains. *FEMS Microbiol. Rev.* **22**, 1–20 (1998).

40. Shah, M. *et al.* A phage-encoded anti-activator inhibits quorum sensing in Pseudomonas aeruginosa. *Mol. Cell* **81**, 571-583.e6 (2021).
41. Al-Shayeb, B. *et al.* Clades of huge phages from across Earth's ecosystems. *Nature* **578**, 425–431 (2020).
42. Sonnhammer, E., Eddy, S. R., Birney, E., Bateman, A. & Durbin, R. Pfam: multiple sequence alignments and HMM-profiles of protein domains. *Nucleic Acids Res.* **26**, 320–322 (1998).
43. Wilson, D. *et al.* SUPERFAMILY - Sophisticated comparative genomics, data mining, visualization and phylogeny. *Nucleic Acids Res.* **37**, D380 (2009).
44. Zdobnov, E. M. & Apweiler, R. InterProScan - An integration platform for the signature-recognition methods in InterPro. *Bioinformatics* **17**, 847–848 (2001).
45. Gumerov, V. M. & Zhulin, I. B. TREND: a platform for exploring protein function in prokaryotes based on phylogenetic, domain architecture and gene neighborhood analyses. *Nucleic Acids Res.* **48**, W72–W76 (2020).
46. Price, M. N., Dehal, P. S. & Arkin, A. P. FastTree 2 - Approximately maximum-likelihood trees for large alignments. *PLoS One* **5**, (2010).
47. Katoh, K., Misawa, K., Kuma, K. I. & Miyata, T. MAFFT: A novel method for rapid multiple sequence alignment based on fast Fourier transform. *Nucleic Acids Res.* **30**, 3059–3066 (2002).
48. Taylor, W. R. Residual colours: a proposal for aminochromography. *Protein Eng. Des. Sel.* **10**, 743–746 (1997).
49. Nguyen, L. T., Schmidt, H. A., Von Haeseler, A. & Minh, B. Q. IQ-TREE: A fast and effective stochastic algorithm for estimating maximum-likelihood phylogenies. *Mol. Biol. Evol.* **32**, 268–274 (2015).
50. Kalyaanamoorthy, S., Minh, B. Q., Wong, T. K. F., Von Haeseler, A. & Jermiin, L. S. ModelFinder: Fast model selection for accurate phylogenetic estimates. *Nat. Methods* **14**, 587–589 (2017).
51. Hoang, D. T., Chernomor, O., von Haeseler, A., Minh, B. Q. & Vinh, L. S. UFBoot2: Improving the Ultrafast Bootstrap Approximation. *Mol. Biol. Evol.* **35**, 518–522 (2018).
52. Letunic, I. & Bork, P. Interactive Tree of Life (iTOL) v4: Recent updates and new developments. *Nucleic Acids Res.* **47**, W256–W259 (2019).
53. Altschul, S. F., Gish, W., Miller, W., Myers, E. W. & Lipman, D. J. Basic local alignment search tool. *J. Mol. Biol.* **215**, 403–410 (1990).

**Data availability**
The huge phage genomes can be downloaded from ggkbase.berkeley.edu/hpae_final/organisms. The mobilome metagenomic data can be downloaded from ggkbase.berkeley.edu/phage-plasmid-virus-protein-families/organisms. The code used to analyze the datasets is available at github.com/hjshim/Pro_ATP.

**Acknowledgements**
We thank Raphaël Méheust and Rohan Sachdeva for the bioinformatic support. We thank the Somatic Cell Genome Editing (SCGE) consortium, particularly the genome-editor working group, for discussions. This work was supported by the National Institutes of Health (NIH) Common Fund Program, Somatic Cell Genome Editing, through an award administered by the National Institute of Allergy and Infectious Diseases [1U01AI142817-01]; PIs: [Jennifer A Doudna and Jillian F. Banfield]).

**Competing interests**
The authors declare no competing interests.

# Tables

**Table 1.** Summary statistics of 353 huge phage genomes and associated ATPase proteins.

| Summary Statistics | Genome length (kbp) | Total proteins | GC content (%) | ATPase domains | ATPase proteins | Multiple ATPase proteins |
|---|---|---|---|---|---|---|
| **Mean** | 291 | 397.2 | 39.20 | 14.29 | 12.59 | 1.70 |
| **Standard deviation** | 89 | 160.68 | 8.99 | 6.14 | 5.28 | 1.30 |
| **Min** | 104 | 106 | 21.98 | 3 | 3 | 0 |
| **Max** | 735 | 1144 | 63.98 | 60 | 47 | 13 |

**Table 2.** Percentage of ATPase per protein by category in the mobilome metagenomes.

|  | Virus | Phage | Plasmid | Other |
|---|---|---|---|---|
| **Number of proteins** | 44,243 | 731,372 | 112,101 | 35,695 |
| **Number of ATPases** | 361 | 2,654 | 2,216 | 266 |
| **ATPase/protein** | 0.82% | 0.36% | 1.98% | 0.75% |

**Figure 1.** Phage life cycle with a list of proteins having an ATPase domain (highlighted in red), potentially involved in the energy consuming activities of each stage.

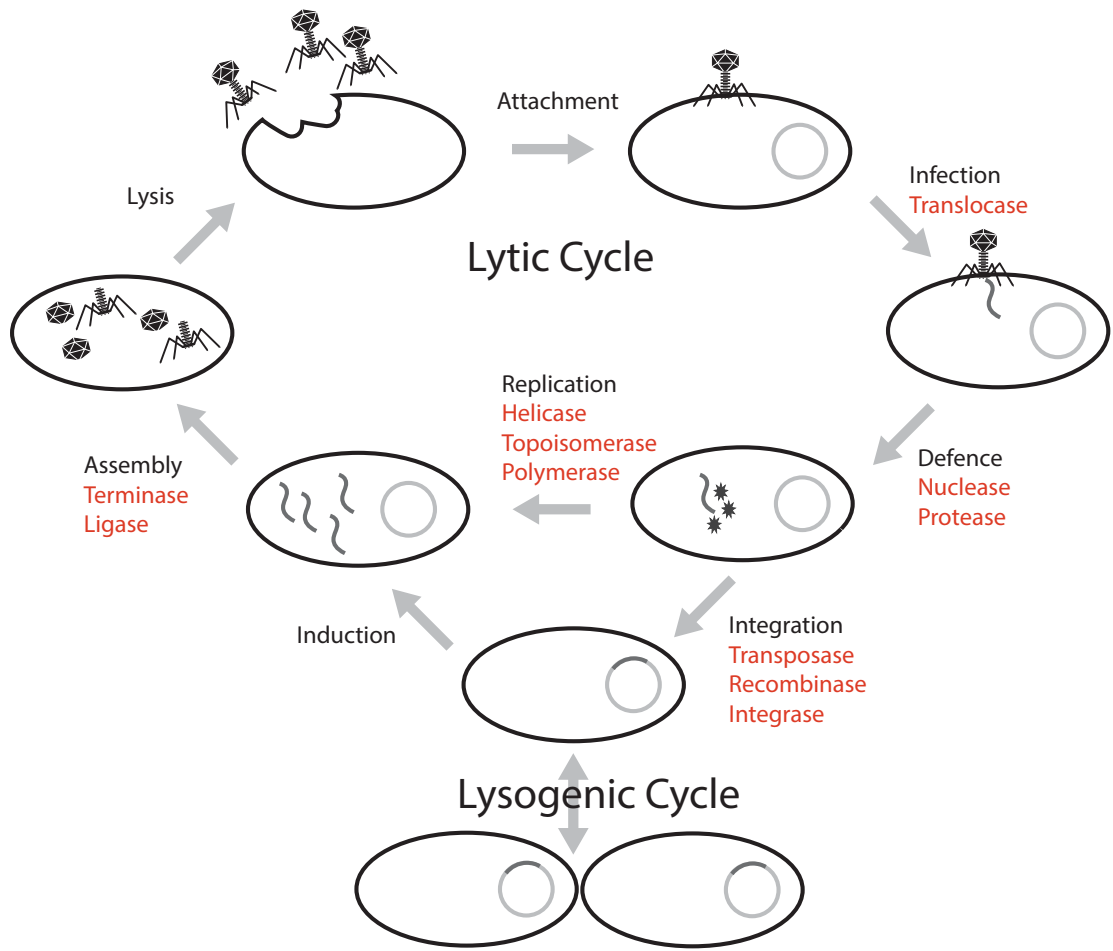

**Figure 2.** Survey of ATPase proteins on the complete genome of a huge phage (PHAGE-A1--js4906-22-3_S10_HugePhage_Circular_26_80_closed).

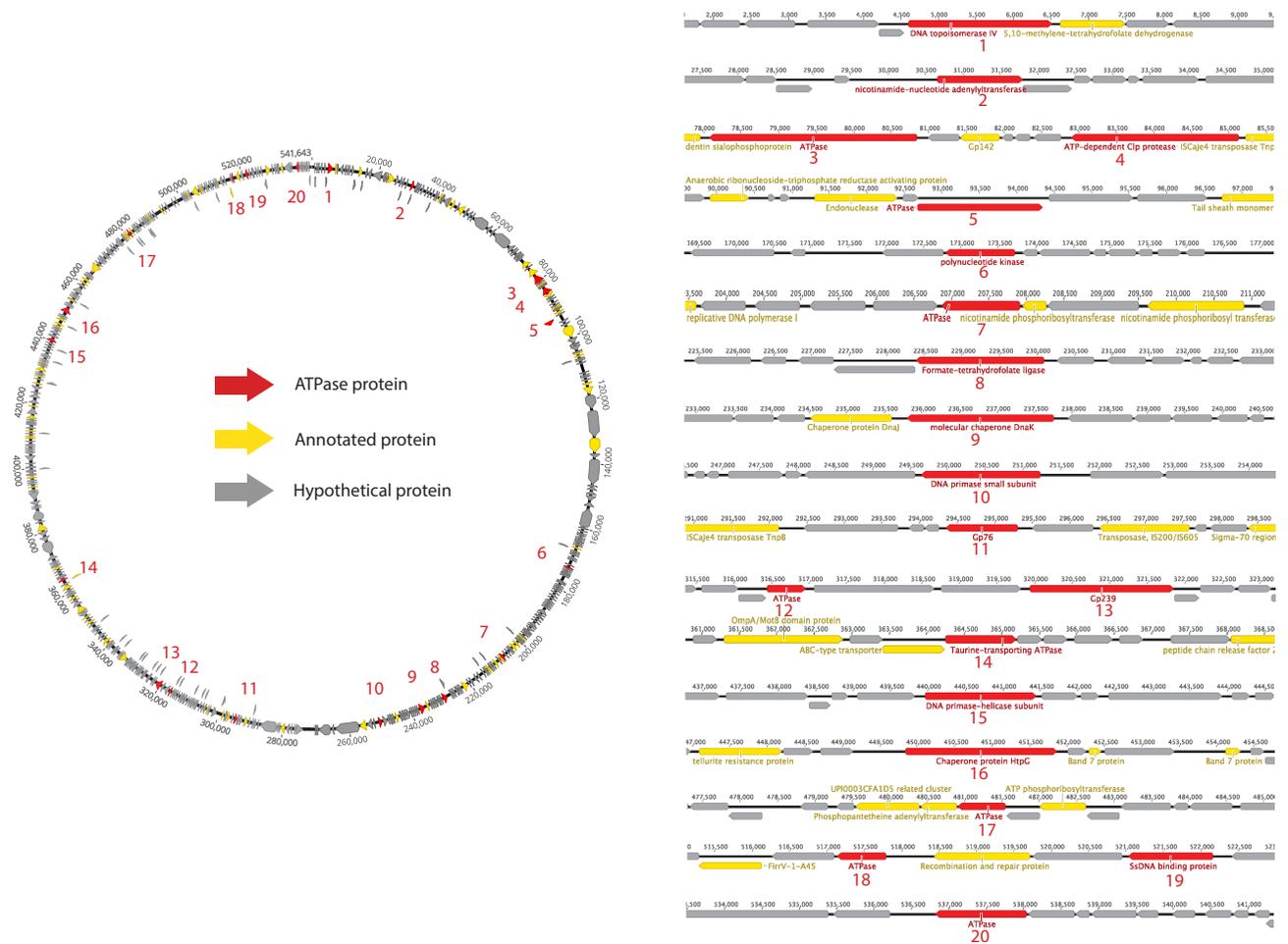

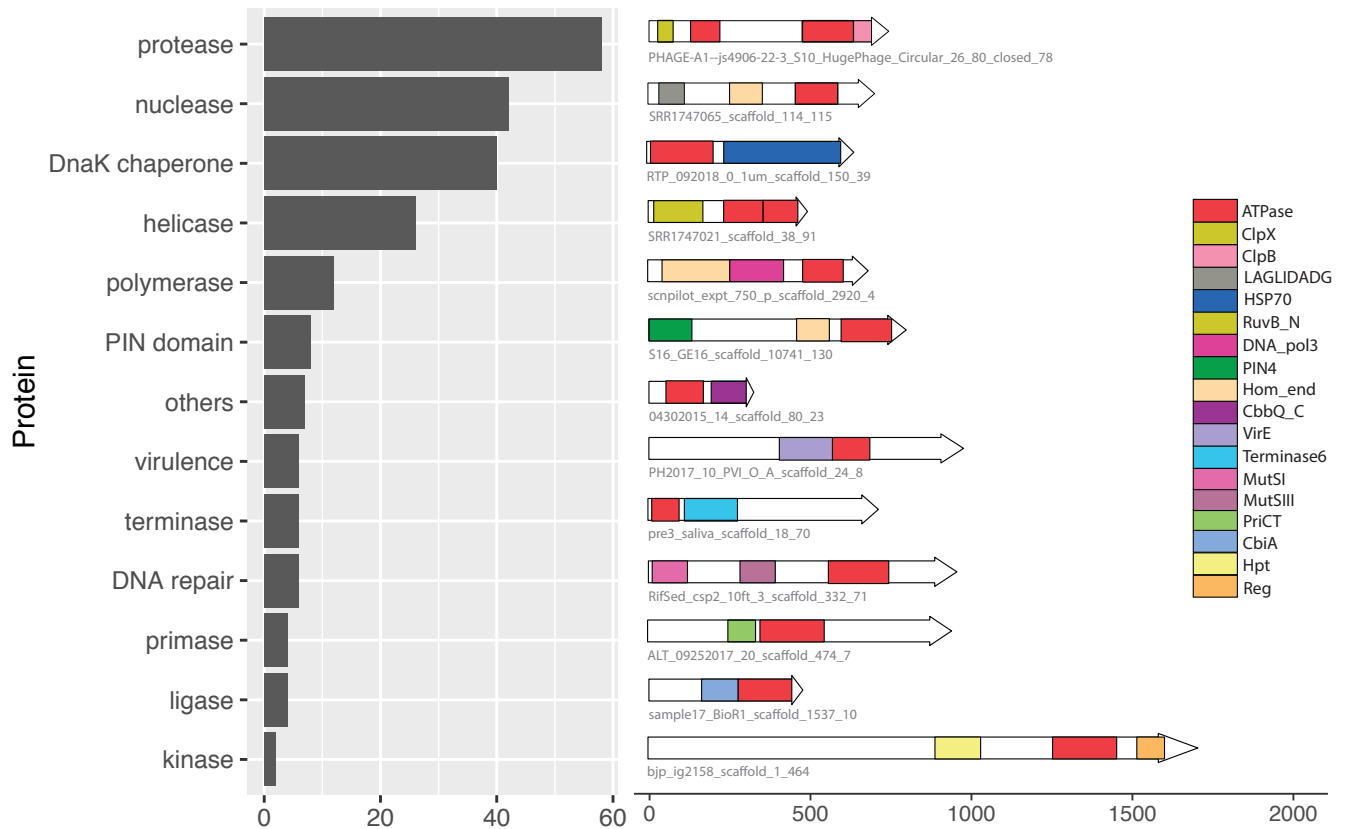

**Figure 3.** Analysis of the functional domains of composite ATPase proteins from the mobilome metagenomes. The bar chart shows the predicted functions of the 221 composite ATPase proteins are nucleic acid and amino acid interacting enzymes. For each category, a representative example of a composite ATPase protein is shown, with all the known domains drawn to scale. The red rectangles indicate the domains predicted to have energy-utilizing functions such as AAA+ and ABC.

**Figure 4.** (A) Multiple sequence alignment of the ATPase domains of all the composite ATPase proteins reveals the conserved regions of the ATPase family (Walker A motif and Walker B motif), as well as other regions (Sensor-1 and Sensor-2). Each ATPase domain is labelled with the biological or enzymatic activity of the functional domain that is fused together as a composite protein. (B) A phylogenetic tree of the ATPase domains from diverse environments is built using the multiple sequence alignment and the structural reconstruction is shown for each class of ATPase domains.

**A**

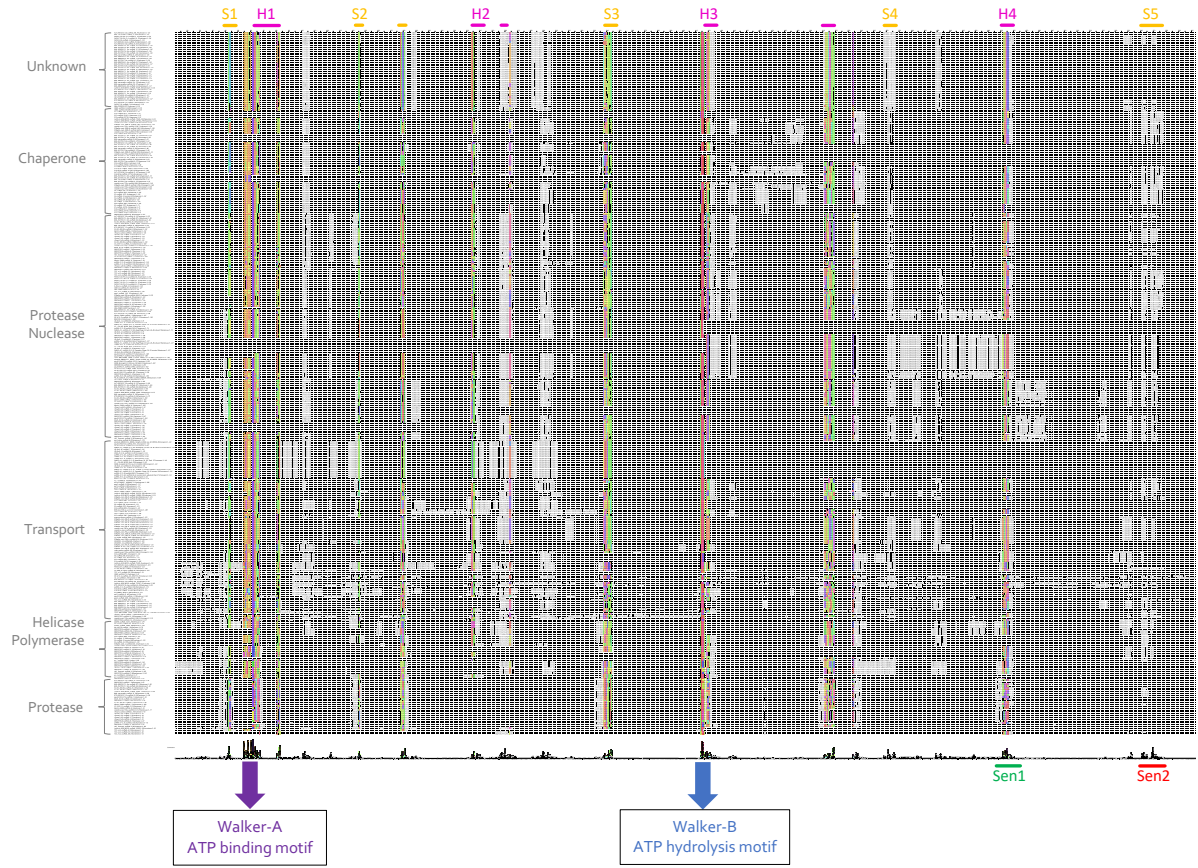

**B**

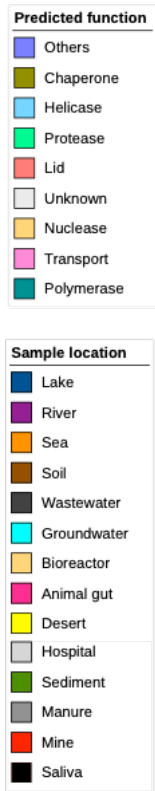
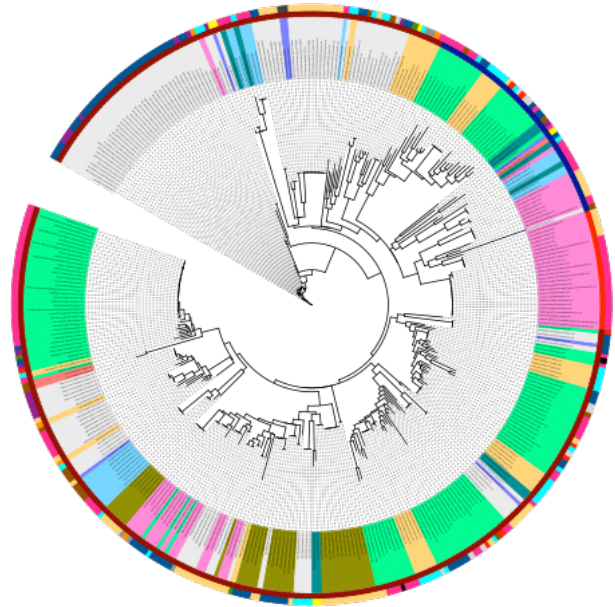
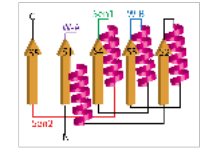
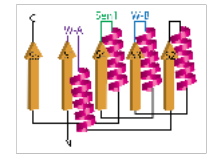
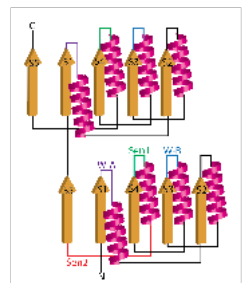

**Figure 5.** (A) Heatmap of genomic neighborhood of composite ATPase proteins, with the x-axis showing the proteins localized ±3 of each composite ATPase protein in the middle. The color bar represents the number of blast hits in the ggKbase database (with the threshold e-value of 1e-30) clustered by similarity. (B) Examples of genomic neighborhood of composite ATPase proteins, with the proteins localized ±3 of each composite ATPase protein indicated in the middle. Most proteins are unannotated and hypothetical proteins, except those with labels. Identical to the genomic neighborhood heatmap, the color represents the number of blast hits in the ggKbase database. Composite ATPase proteins that are found to have highly conserved operons (indicated with green stars in 5(A)). (C) Composite ATPase proteins that are found to have highly conserved pairs with a sequence overlap (indicated with blue stars in 5(A)).

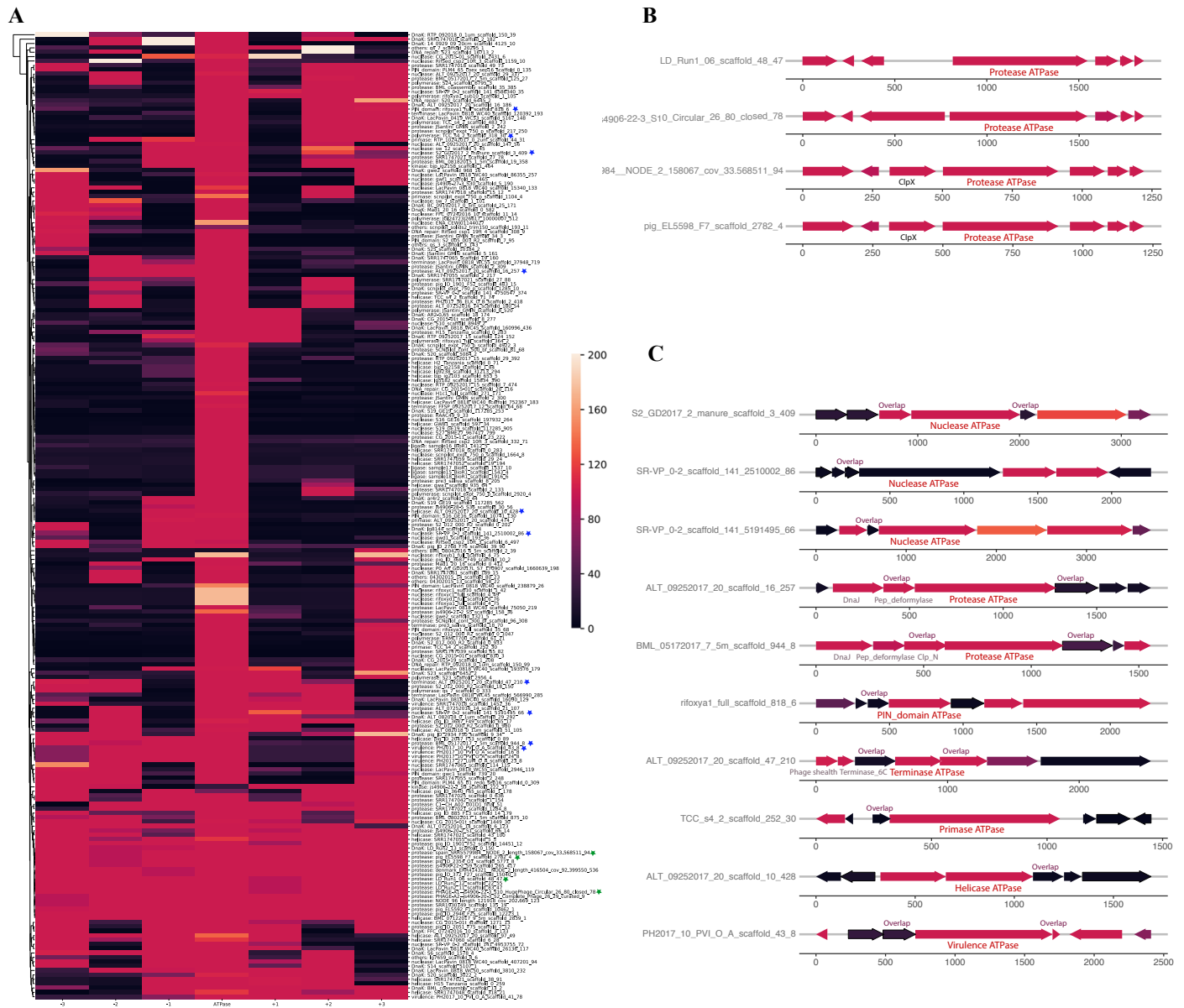